# Calculation of the Critical Temperature for 2- and 3-Dimensional Ising Models and for 2-Dimensional Potts Models Using the Transfer Matrix Method

## M. Ghaemi,*,† G. A. Parsafar,† and M. Ashrafizaadeh‡

*Department of Chemistry, Department of Mechanical Engineering, Isfahan University of Technology, Isfahan 84154, Iran*

A new graphical method is developed to calculate the critical temperature of 2- and 3-dimensional Ising models as well as that of the 2-dimensional Potts models. This method is based on the transfer matrix method and using the limited lattice for the calculation. The reduced internal energy per site has been accurately calculated for different 2-D Ising and Potts models using different size-limited lattices. All calculated energies intersect at a single point when plotted versus the reduced temperature. The reduced temperature at the intersection is 0.4407, 0.2746, and 0.6585 for the square, triangular, and honeycombs Ising lattices and 1.0050, 0.6309, and 1.4848 for the square, triangular, and honeycombs Potts lattices, respectively. These values are exactly the same as the critical temperatures reported in the literature, except for the honeycomb Potts lattice. For the two-dimensional Ising model, we have shown that the existence of such an intersection point is due to the duality relation. The method is then extended to the simple cubic Ising model, in which the intersection point is found to be dependent on the lattice sizes. We have found a linear relation between the lattice size and the intersection point. This relation is used to obtain the critical temperature of the unlimited simple cubic lattice. The obtained result, 0.221(2), is in a good agreement with the accurate value of 0.22165 reported by others.

## Introduction

For many years, the lattice statistics has been the subject of intense research interests. Although, at zero magnetic field, there is an exact solution for the 2-dimensional (2-D) Ising model,[1] however, there is no such a solution for the 3-dimensional (3-D) Ising and 2-D Potts models. The Potts models are the general extension of the Ising model with $q$-state spin lattice,[2] i.e., the Potts model with $q = 2$ is equivalent to the Ising model. In the absence of an exact solution, series expansions remain as one of the most useful tools in the investigation of the critical properties of a model system. Domb[3,4] has provided a historical review of series expansions solutions for the Ising model. A similar review article for the Potts model is provided by Wu[5] and more recently by Biggs.[6,7] In the series expansions method, which is based on the graph theory,[7] the partition function is expanded as a series that counts closed graphs of all possible number of bonds of the lattice. In addition to the complexity of the calculation, the disadvantage of this method is that the precision of the calculated critical data depends on the number of terms in the truncated series.

Other techniques, which have been developed during the past two decades, are the simulation methods,[8–12] especially the Monte Carlo method. The precision of the calculated critical point obtained by this method depends on the number of particles used in the model system i.e., the lattice size. Because of the limitation in computer resources we cannot freely increase the size of lattice.

In a recent work, Ranjbar and Parsafar[13] used a limited number of rows, but with the same coordination number for each site to set up the transfer matrix for the square lattice in the Ising model. The resultant matrix was solved for the model with the number of rows equal to or less than 10, from which the exact thermodynamic properties were obtained, in the absence of any magnetic field. The calculated reduced internal energy per site, $u(K)$, was then plotted versus the reduced temperature, $K$, for the square lattice with different lattice sizes, where $K = j/kT$ and $j > 0$ is the coupling energy for the nearest-neighbor spins. It was observed that all of them intersect at a single point known as the critical temperature. Since the order of the transfer matrix becomes too large for both the Potts and the 3-D Ising models, the method cannot be extended to calculate the critical temperature for these models, analytically.

In the present work we have extended the method of Ranjbar and Parsafar[13] to numerically calculate the critical temperature for the 2-D Potts and Ising and 3-D Ising models. In the following section, we have shown that the existence of the duality relation for the two-dimensional Ising model implies that the $u(K)$ at the critical temperature must be independent of the size of lattice. Owing to this fact, we have been able to calculate the critical temperature of the square, triangular, and honeycomb lattices. We have then extended the method to the 3-state Potts models for the 2-D lattices. We have shown that our method can be easily used to obtain the critical point for the 2-D Potts models. The method has been also extended to the 3-D Ising model. Unlike the 2-D models, according to our results, the location of the intersection point depends on the lattice sizes, in other words there is no single intersection point. We have found that the location of the intersection point versus the lattice size is almost linear. We have extrapolated the line to the unlimited lattice size to obtain the critical temperature of simple cubic lattice.

## The Duality Relation and The Common Intersection Point

Consider a square lattice with the periodic boundary condition composed of slices, each with $p$ rows, where each row has $r$ sites. Each slice has then $p \cdot r = N$ sites and the coordination number of all sites is the same. In the 2-D Ising model, for any site we define a spin variable $\sigma(i,j) = \pm 1$, in such a way that $i = 1,...,r$ and $j = 1,...,p$. We include the periodic boundary condition as:

$$\sigma(i + r, j) = \sigma(i, j) \quad (1)$$

$$\sigma(i, j + p) = \sigma(i, j) \quad (2)$$

---

* Corresponding author, email: ghaemi@saba.tmu.ac.ir

† Department of Chemistry.
‡ Department of Mechanical Engineering.

We take only the interactions among the nearest neighbors into account. The configurational energy for the model is given as,

$$E(\sigma) = -j\sum_{i=1}^{r}\sum_{j=1}^{p}\{\sigma(i,j)\sigma(i+1,j) + \sigma(i,j)\sigma(i,j+1)\} \quad (3)$$

The canonical partition function, $Z(K)$, is

$$Z(K) = \sum_{\{\sigma\}} e - \frac{E(\sigma)}{kT} \quad (4)$$

Substitution of eq 3 into eq 4 gives,[3]

$$Z(K) = \sum_{\sigma(\{i\},1)} ... \sum_{\sigma(\{i\},p)} \langle 1|B|2\rangle\langle 2|B|3\rangle...\langle p|B|1\rangle \quad (5)$$

where

$$|j\rangle = |\sigma(1,j)\rangle \otimes |\sigma(2,j)\rangle... \otimes |\sigma(r,j)\rangle \quad (6)$$

and

$$\sum_{\sigma(\{i\},j)} = \sum_{\sigma(1,j)}\sum_{\sigma(2,j)}...\sum_{\sigma(r,j)} \quad (7)$$

The element $B_{t,t+1}$ of the transfer matrix $B$ is defined as,

$$B_{t,t+1} = \langle t|B|t+1\rangle = \exp\{K\sum_{i=1}^{r}[\sigma(i,t)\sigma(i,t+1) + \sigma(i,t)\sigma(i+1,t)]\} \quad (8)$$

By orthogonal transformation, the $B$ matrix can be diagonalized, where eq 4 for the large values of $p$ can be written as[13]

$$Z(K) = (\lambda_{max})^p \quad (9)$$

where the $\lambda_{max}$ is the largest eigenvalue of $B$. From the well-known thermodynamic relation, $A = -kT\ln Z$, along with eq 8 the following results can be obtained:[13]

$$a(K) = -\frac{A}{NkT} = \frac{\ln \lambda_{max}}{r} \quad (10)$$

$$u(K) = \frac{-E}{Nj} = \frac{\partial a(K)}{\partial K} \quad (11)$$

where $u(K)$ and $a(K)$ are the reduced internal energy and Helmholtz free energy per site, respectively. The value of $\lambda_{max}$ as a function of $K$ was given by Kaufman[14] for the limited lattice as:

$$\frac{\ln \lambda_{max}}{r} = \frac{1}{2r}\{\gamma_1 + \gamma_2 + \gamma_3 + ... + \gamma_{2r-1}\} + \frac{1}{2}\ln(2\sinh 2K) \quad (12)$$

where

$$\cosh(\gamma_i) = \cosh(2\tilde{K})\cosh(2K) - \sinh(2\tilde{K})\sinh(2K)\cos\left(\frac{i\pi}{r}\right) \quad (13)$$

and $\tilde{K}$ is the reduced temperature of the dual lattice. By the well-known duality relation[4,5,14], i.e., $e^{-2\tilde{K}} = \tanh(K)$, eq 13 can be written as:

$$\cosh(\gamma_i) = \coth(2K)\cosh(2K) - \cos\left(\frac{i\pi}{r}\right) \quad (14)$$

Substitution of eqs 10 and 12 into eq 11 gives,

$$u(K) = \frac{1}{2}\frac{\partial}{\partial K}\ln(\sinh(2K)) + \frac{1}{2r}\frac{\partial}{\partial K}\{\gamma_1 + \gamma_2 + ... + \gamma_{2r-1}\} \quad (15)$$

Differentiation of eq 14 with respect to $K$ gives,

$$\frac{\partial}{\partial K}\cosh(\gamma_i) = \frac{\partial\cosh(\gamma_i)}{\partial\gamma_i}\frac{\partial\gamma_i}{\partial K} = 2\cosh(2K)(1 - 1/\sinh^2(2K)) \quad (16)$$

Because at the critical temperature $\sinh^2(2K_c) = 1$ (see ref 14), we may conclude from eq 16 that $\partial\gamma_i/\partial K = 0$ and, hence, $u(K_c)$ is independent of the lattice size $r$. The same conclusion can be easily obtained for both the triangular and honeycomb Ising lattices, see appendix A.

The numerical calculation of $u(K)$ can be easily programmed by the well-known mathematical softwares such as Maple, Mathlab, Mathematica,..., and ARPACK. For the square lattice with size r, by using eq 8, the elements of the $B$ matrix have been calculated numerically. We have used ARPACK to calculate the largest eigenvalue, $\lambda_{max}$, of the $B$ matrix for different $K$ values, and along with eqs 10 and 11 the values of $u(K)$ have been calculated. The calculation has been repeated for different lattice sizes. Then, by drawing $u(K)$ versus $K$ for different lattice sizes, as expected, all of them intersect at a single point known as the critical point. Such a calculation was also been done for the triangular and honeycomb lattices, see Figures 1 and 2. The results of such calculations are compared with the exact values[3,4] in Table 1, which are exactly the same.

**The Potts Model**

Although we do not know the exact solution of the Potts model for the 2-dimension at present time, a large amount of numerical information has been accumulated for the critical properties of the various Potts models. For further information, see the excellent review written by Wu[5] or the references given by him. The reason for the extension of our approach to the Potts model is the fact that such a model is an important testing ground for different methods and approaches in the study of critical point theory. Although in the absence of any exact solution for the Potts model we cannot analytically extend the duality argument to drive the size independence of the $u(K_c)$, we may test our approach numerically.

Consider a square lattice with the periodic boundary condition composed of slices, each with $p$ rows, where each row has $r$ sites. Then each slice has $p \cdot r = N$ sites and the coordination number of all sites are the same. For any site we define a spin variable $\sigma(i, j) = 0, \pm 1$ so that $i=1,...,r$ and $j=1,...,p$. The configurational energy of the standard 3-state Potts model is given as,[5]

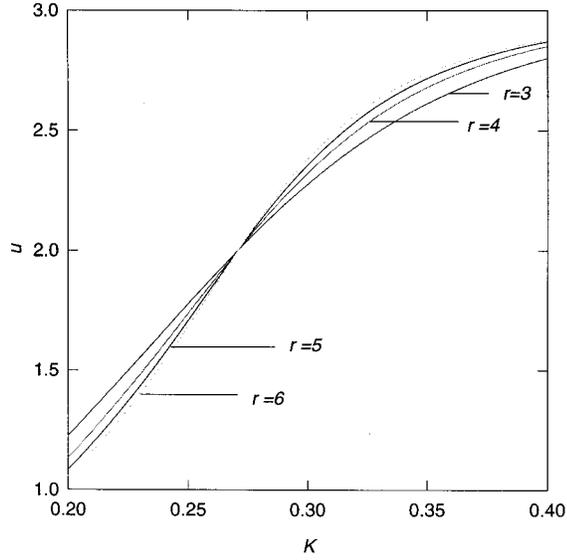

**Figure 1.** The reduced internal energy versus $K$ for triangular lattice of the Ising model.

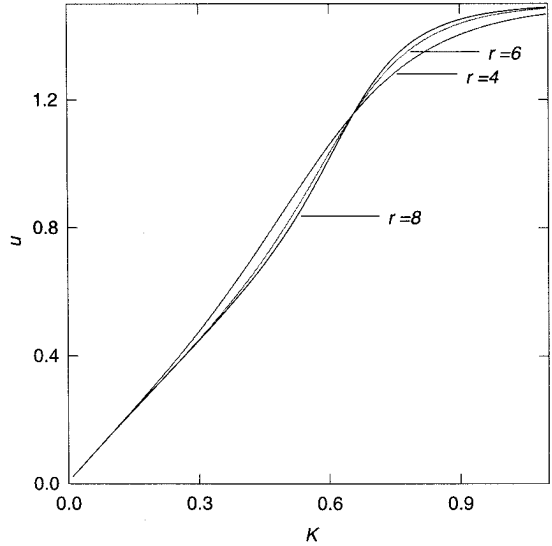

**Figure 2.** Same as Figure 1 for honeycomb lattice of the Ising model.

**TABLE 1: The Calculated Critical Temperatures Compared to Those Reported by Others, for Different Given Ising and Potts Models**

| lattices | Potts models Our results | Other methods | Ising models Our results | Other methods |
|---|---|---|---|---|
| Square | $1.0050 \pm 0.0001$ | $1.005052...$ [c] | $0.4407 \pm 0.0001$ | $0.440687...$ [a] |
| Triangular | $0.6309 \pm 0.0001$ | $0.630944...$ [c] | $0.2746 \pm 0.0001$ | $0.274653...$ [a] |
| Honeycomb | $1.4848 \pm 0.0001$ | $1.484208...$ [c] | $0.6585 \pm 0.0001$ | $0.658478...$ [a] |
| Simple cubic | | | $0.2212 \pm 0.0001$ | $0.2216(5)$ [b] |

[a] Exact value given in refs 3, 4, and 20. [b] From the recent simulation methods given in refs 11 and 12. [c] Computed values from the given formula in ref 5.

$$E(\sigma) = \sum_{i=1}^{r}\sum_{j=1}^{p} -j\{\delta_{\sigma(i,j),\sigma(i,j+1)} + \delta_{\sigma(i,j),\sigma(i+1,j)}\} \quad (17)$$

where

$$\delta_{i,j} = 1 \text{ for } i = j$$
$$\delta_{i,j} = 0 \text{ for } i \neq j \quad (18)$$

The canonical partition function, $Z(K)$, is

$$Z(K) = \sum_{\sigma(\{i\},1)} ... \sum_{\sigma(\{i\},p)} \langle 1|C|2\rangle\langle 2|C|3\rangle...\langle p|C|1\rangle \quad (19)$$

The element $C_{t,t+1}$ of the transfer matrix $C$ is defined as,

$$\langle t|C|t+1\rangle = \exp\{K\sum_{i=1}^{r}[\delta_{\sigma(i,t),\sigma(i+1,t)} + \delta_{\sigma(i,t),\sigma(i,t+1)}]\} \quad (20)$$

The value of $u(K)$ was calculated, numerically, and drawn versus $K$ for different lattice sizes. Again, all of them intersect at a single point, see Figure 3, which is the critical point. Such a calculation was also carried out for the honeycomb and triangular lattices, for which the matrix elements are given in appendix B, along with those for the Ising models. For each case we have observed a common intersection point which is given in Table 1.

### 3-Dimensional Ising Model

Although there is not any well-known duality relation in the three-dimensional Ising model,[15] our method can easily be extended to the 3-dimensional lattice. For simplicity, consider a simple cubic lattice with the periodic boundary condition composed of slices, each has $p$ layers. Each layer has $r$ rows and each row has $m$ sites. Then, each slice (limited lattice) has $p \cdot r \cdot m = N$ sites. A layer of a slice with its nearest neighbor slices is shown in Figure 4 in which the periodic boundary condition is taken into account. For any site we define a spin variable $\sigma(i, j, k) = \pm 1$ in such a way that $i=1,...,m$, $j=1,...,r$, $l=1,...,p$. We include the periodic boundary condition as:

$$\sigma(i + m, j, l) = \sigma(i, j, l) \quad (21)$$

$$\sigma(i, j + r, l) = \sigma(i, j, l) \quad (22)$$

$$\sigma(i, j, l + p) = \sigma(i, j, l) \quad (23)$$

The configurational energy is given as,

$$E(\sigma) = -j\sum_{i=1}^{m}\sum_{j=1}^{r}\sum_{l=1}^{p}\{\sigma(i,j,l)\sigma(i+1,j,l) + \sigma(i,j,l)\sigma(i,j+1,l) + \sigma(i,j,l)\sigma(i,j,l+1)\} \quad (24)$$

The canonical partition function, $Z(K)$, is

$$Z(K) = \sum_{\sigma(\{i\},\{j\},1)} ... \sum_{\sigma(\{i\},\{j\},p)} \langle 1|D|2\rangle\langle 2|D|3\rangle...\langle p|D|1\rangle \quad (25)$$

where

$$|l\rangle = |\sigma(1,1,l)\rangle \otimes |\sigma(1,2,l)\rangle ... \otimes |\sigma(m,r,l)\rangle \quad (26)$$

and

$$\sum_{\sigma(\{i\},\{j\},l)} = \sum_{\sigma(1,1,l)}\sum_{\sigma(1,2,l)} ... \sum_{\sigma(m,r,l)} \quad (27)$$

where the element $D_{l,l+1}$ of the transfer matrix $D$ is defined as,

$$D_{l,l+1} = \langle l|D|l+1\rangle = \exp\{K\sum_{i=1}^{m}\sum_{j=1}^{r}[\sigma(i,j,l)\sigma(i+1,j,l) + \sigma(i,j,l)\sigma(i,j+1,l) + \sigma(i,j,l)\sigma(i,j,l+1)]\} \quad (28)$$

From the diagonalization of the $D$ matrix, $u(K)$ and $a(K)$ can be numerically calculated. The PARPACK package is used to diagonalize the matrix with the order of $2^{mr}$, from which the $\lambda_{max}$ was calculated with a high precision. All parallel computa-tions are performed on a PC cluster composed of four dual computers with PIII 550 MHz processors. The calculated reduced energy for simple cubic lattice with ($r = 3$, $m = 3$), ($r = 4$, $m = 4$), and ($r = 4$, $m = 4$) was plotted vesus $K$, shown in Figure 5. As shown in Figure 5, the location of the intersection point depends on the sizes of the lattices and it may be related to the fact that there is not any well-known duality relation for the 3-dimensional Ising model.[15] As shown in Figure 5, with increasing the sizes of the lattices the value of $K = j/kT$ at the intersection point decreases. To obtain the value of $K_C$, we have to obtain the intersection point for two unlimited lattices with different sizes, however, such a point may be predicted if we have an expression for the intersection point in terms of $1/n$, where $n = mrm'r'$. To do so, the intersection points of Figure 5 have been plotted versus $1/n$. As shown in Figure 6, the points are located almost on the straight line of $K_c = 2.4454/n + 0.2212$ with $R^2 = 0.99997$, where $R^2$ is the linear correlation. The value of $K_c = 0.2212$ for the intercept is the critical temperature for the unlimited simple cubic lattice of Ising model.

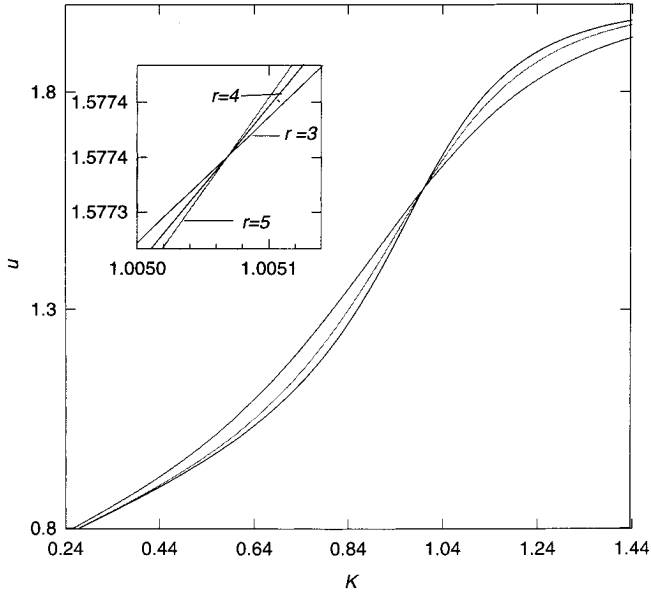

**Figure 3.** Same as Figure 1 for the square lattice of the Potts model.

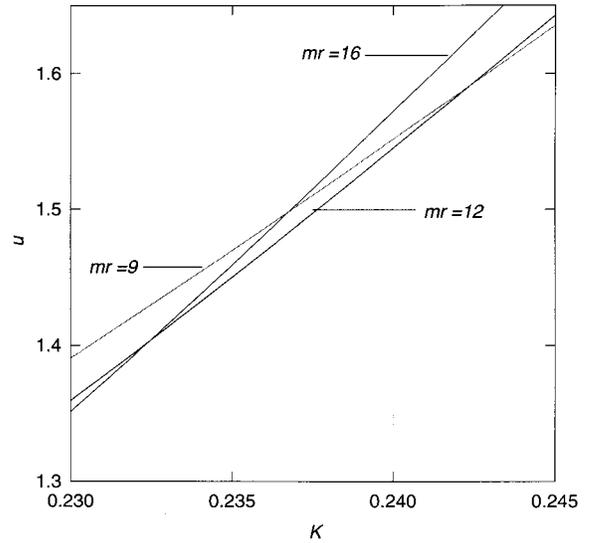

**Figure 5.** Same as Figure 1 for simple cubic lattice of Ising model.

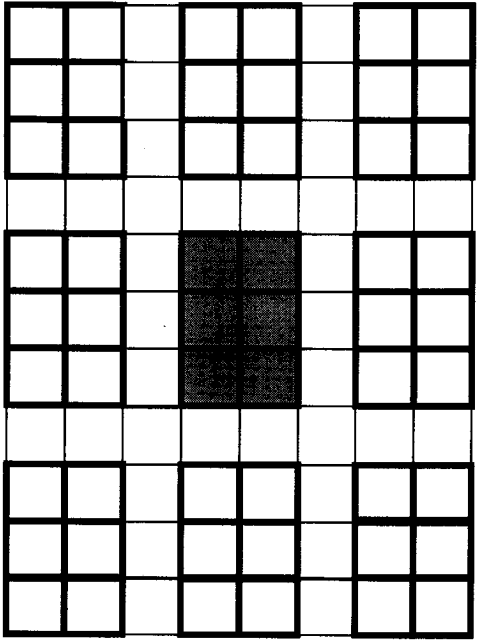

**Figure 4.** A layer of a slice (at the center) with its nearest neighbors, with $r = 3$ and $m = 4$.

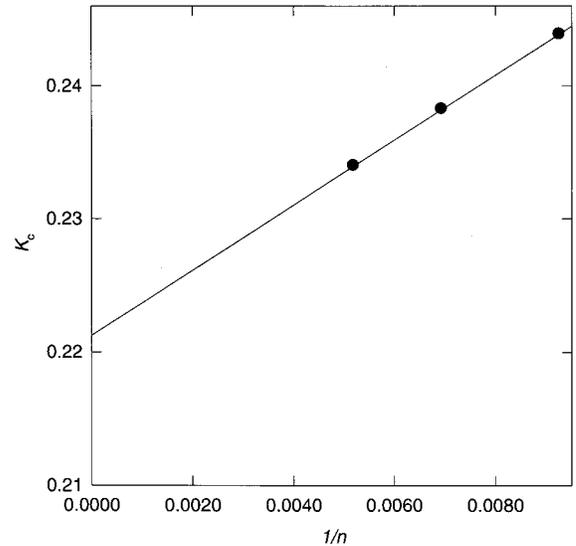

**Figure 6.** Using the intersection points of Figure 5 to obtain the critical temperature of the 3-D Ising model.

### Conclusion

We have shown that our graphical method, in principle, can be applied to the 2-D and 3-D Ising and Potts models. For the 2-D Ising models, the existence of a duality relation can be used to accurately calculate the critical temperature. In fact, our numerical results are exactly the same as the reported exact values,[3,4] for such cases.

The only exact solution of the Potts model known to date is the Onsager's solution[1] of the two-state Potts model, and the transformation to the dual is known analytically[1,14] for this case. For the other models, the duality relation is expected to be at work, but its existence has only been tested numerically.[15,16] The existence of the conjectured duality relation[3,5,6,18,19] has been used to calculate the critical temperature. Although our calculated critical temperatures for the square and triangular lattices are exactly the same as those predicted values by other methods,[5] for the honeycomb lattice there is a small difference between our calculated value and that obtained from the formula given by Wu[5] (i.e., Equation 5.1b of ref 5). For this reason, such difference for the honeycomb lattice may be attributed to the incompleteness of eq 5.1b of ref 5.

We have shown that our method can be used to calculate the critical temperature of the simple cubic Ising model. The obtained result, 0.221(2), is in a good agreement with the accurate value of 0.22165, obtained from the recent simulation methods.[11,12] The small inaccuracy in our results is due to the limitation dictated by available computer resources. However, we expect to obtain more accuracy if matrixes of higher orders than $2^{16}$ are used.

**Acknowledgment.** We acknowledge the Iranian National Research Council for the financial support, and also Dr. B. Mirza for his useful comments.

**Appendix A**

Based on the duality argument, the value of $\lambda_{max}$ as a function of $K$ was given by Wannier[20] for the limited triangular lattice as,

$$\frac{\ln \lambda_{max}}{r} = \frac{1}{2}\ln(2\sinh 2K) + \frac{1}{2r}\{\gamma_1^{(1)} + \gamma_2^{(1)} + \gamma_1^{(2)} + \gamma_2^{(2)} + \dots + \gamma_1^{(r/2)} + \gamma_2^{(r/2)}\} \quad (A.1)$$

and for it's dual (i.e., the honeycomb) lattice, it can be shown that,

$$\frac{\ln \lambda^*_{max}}{r} = \ln(2\sinh 2K^*) + \frac{1}{2r}\{\gamma_1^{(1)} + \gamma_2^{(1)} + \gamma_1^{(2)} + \gamma_2^{(2)} + \dots + \gamma_1^{(r/2)} + \gamma_2^{(r/2)}\} \quad (A.2)$$

where $K^*$ and $K$ are related by the duality relation of $\sinh 2K^* \sinh 2K = 1$. The expressions for $\gamma_1^{(i)}$ and $\gamma_2^{(i)}$ are given as,

$$\cosh\left(\frac{\gamma_1^{(i)}}{2}\right) = \frac{1}{2}\left[\kappa^{-1} + \cos^2\frac{\omega(i)}{2}\right]^{1/2} + \frac{1}{2}\cos\frac{\omega(i)}{2} \quad (A.3)$$

and

$$\cosh\left(\frac{\gamma_2^{(i)}}{2}\right) = \frac{1}{2}\left[\kappa^{-1} + \cos^2\frac{\omega(i)}{2}\right]^{1/2} - \frac{1}{2}\cos\frac{\omega(i)}{2} \quad (A.4)$$

where

$$\omega(i) = \frac{4\pi i}{r} \quad (A.5)$$

and

$$\kappa^{-1} = \frac{(e^{4K} + 1)^2}{(e^{4K} - 1)} \quad (A.6)$$

Substitution of eqs 10 and A.1 into eq 11 gives,

$$u(K) = \frac{1}{2}\frac{\partial}{\partial K}\ln(\sinh(2K)) + \frac{1}{2r}\frac{\partial}{\partial K}\{\gamma_1^{(i)} + \gamma_2^{(i)} + \dots + \gamma_2^{(r/2)}\} \quad (A.7)$$

Differentiation of eq A.3 or A.4 with respect to $K$ gives,

$$\frac{\partial}{\partial K}\cosh(\gamma_{1(2)}^{(i)}/2) = \frac{\partial\cosh(\gamma_{1(2)}^{(i)}/2)}{\partial\gamma_{1(2)}^{(i)}}\frac{\partial\gamma_{1(2)}^{(i)}}{\partial K} =$$
$$\frac{1}{4}\left[\kappa^{-1} + \cos^2\frac{\omega(i)}{2}\right]^{-1/2}\frac{\partial\kappa^{-1}}{\partial K} \quad (A.8)$$

Because at the critical temperature $\exp(2K_C) = \sqrt{3}$ (see ref 20), we may conclude from eq A.8 that $\partial\gamma_{1(2)}^{(i)}/\partial K = 0$ at this point, and hence $u(K_c)$ is independent of the lattice size, $r$. The same result can easily be obtained for the honeycomb lattice, as well.

**Appendix B**

The elements of the transfer matrix $B$ for the triangular Ising lattice, $B_{j,j+1}$,

$$B_{j,j+1} = \exp\{K\sum_{i=1}^{r}[\sigma(i,j)\sigma(i,j+1) + \sigma(i,j)\sigma(i+1,j+1) + \sigma(i,j)\sigma(i+1,j)]\} \quad (B.1)$$

honeycomb Ising lattice, $H_{j,j+1}$, with even number for $r$,

$$H_{j,j+1} = \exp\{K[\sum_{i=1}^{r}\sigma(i,j)\sigma(i+1,j) + \sum_{i=1}^{(r/2)}\sigma(2i-1,j)\sigma(2i,j+1)]\} \quad (B.2)$$

triangular Potts lattice, $C_{j,j+1}$,

$$C_{j,j+1} = \exp\{K\sum_{i=1}^{r}[\delta_{\sigma(i,j),\sigma(i,j+1)} + \delta_{\sigma(i,j),\sigma(i+1,j+1)} + \delta_{\sigma(i,j),\sigma(i+1,j)}]\} \quad (B.3)$$

and honeycomb Potts lattice, $M_{j,j+1}$, with even number for $r$:

$$M_{j,j+1} = \exp\{K[\sum_{i=1}^{r}\delta_{\sigma(i,j),\sigma(i+1,j)} + \sum_{i=1}^{r/2}\delta_{\sigma(2i-1,j),\sigma(2i,j+1)}]\} \quad (B.4)$$

**References and Notes**

(1) Onsager, L. *Phys. Rev.* **1944**, *65*, 117.
(2) Potts, R. B. *Proc. Camb. Philos. Soc.* **1952**, *48*, 106.
(3) *The Critical Point. The Historical Introduction to The Modern Theory of Critical Phenomena*; Domb, C. Ed.; Tailor & Francis: New York, 1996.
(4) *Phase Transitions and Critical Phenomena*; Domb, C.; Green, M. S. Eds.; Vol. 3; Academic Press: New York, 1974.
(5) Wu, F. y. *Rev. Mod. Phys.* **1982**, *54*, 235.
(6) Biggs, N. l.; Shrock, R. *J. Phys. A (Lett.)*, **1999**, *32*, L489.
(7) *Algebraic Graph Theory*; Biggs, N. L. Ed.; Cambridge University Press: Cambridge, 1993.
(8) Picco, M.; Ritort, F. *Physica A*. **1998**, *250*, 46.
(9) Markham, J. F.; Kieu, T. D. *Nucl. Phys. B*. **1998**, *63*, 970.
(10) Machta, J.; Newman, M. E. J.; Chayes, L. B. *Phys. Rev. E*. **2000**, *62*, 8782.
(11) Talapov, A. L.; Blöte, H. W. J. *J. Phys. A. Math. Gen*. **1996**, *29*, 5727.
(12) Butera, P.; Comi, M. *Phys. Rev. B*. **2000**, *62*, 14837.
(13) Ranjbar, Sh. and Parsafar, G. A. *J. Phys. Chem. B*. **1999**, *103*, 7514.
(14) Kaufman, B. *Phys. Rev.* **1949**, *76*, 1232.
(15) Carmona, J. M.; Di Giacomo, A.; Lucini, B. *Phys. Lett. B*. **2000**, *485*, 126.
(16) Di Giacomo, A.; Lucini, B.; Montesi, L.; Paffuti, G. *Phys. Rev. D*. **2000**, *61*, 034504.
(17) Kim, D.; Joseph, R. J. *J. Phys. C*. **1974**, *7*, L167.
(18) Baxter, R. J.; Temperley, H. N. V.; Ashley, S. E. *Proc. R. Soc. London. Ser A*. **1978**, *358*, 535.
(19) Wu, F. Y.; Wang, Y. K. *J. Math. Phys.* **1976**, *17*, 439.
(20) Wannier, G. H. *Phys. Rev.* **1950**, *79*, 357.